\documentclass[12pt]{article}

\usepackage{natbib}
\usepackage{times}
\usepackage{epsf}

\begin{document}

\def\btdplot#1#2{\leavevmode\epsfxsize=#2\textwidth \epsfbox{#1}}

\title{Remote sensing of bubble clouds in seawater}

\author{Piotr J. Flatau, Maria Flatau \\
Scripps Institution of Oceanography, University of California, San
Diego, \\La Jolla, California 92093-0221. e-mail pflatau@ucsd.edu
\\ \\ J. R. V. Zaneveld \\Oregon State University, USA \\ \\
Curtis D. Mobley
\\Sequoia Scientific, Inc., USA \\ \\
In press: 2000, Quarterly Journal of the Royal Meteorological
Society}

\maketitle

\abstract{We report on the influence of submerged bubble clouds on
the remote sensing properties of water. We show that the optical
effect of bubbles on radiative transfer and on the estimate of the
ocean color is significant. We present a global map of the volume
fraction of air in water derived from daily wind speed data. This
map, together with the parameterization of the microphysical
properties, shows the possible significance of bubble clouds on
the albedo of incoming solar energy.}


\section{Introduction}\label{sec:intro} ``The effect of bubbles on
the color of the sea may be observed in breaking waves... Where a
great many bubbles have been entrained by a breaking wave it is
white. But where there are fewer of them it is blue-green or
green, brighter than the sea but not as bright as the foamiest
parts of the wave. Even after a wave has broken and the water is
again quiescent, a pastel green patch often remains, slowly fading
into the surrounding sea as the bubbles dissipate. Thus the effect
of bubbles on the color of the sea is similar to that of solid
particles \citep{Bohren87a}.'' Bubbles within the water and foam
on its surface \citep{Bukata95a,Frouin96a,Stramski84a}  can
predominate in determining the radiative transfer properties of
the sea surface at higher wind speeds. However, there is a limited
knowledge about the radiative transfer properties of bubble
clouds, their inherent optical properties (IOP), and  their global
climatology. \cite{Mobley94a} and \cite{Bukata95a} discuss
qualitatively the surface properties of bubble clouds.
\cite{Frouin96a} performed spectral reflectance measurements of
sea foam at the Scripps Institution of Oceanography Pier. They
observed a decrease of the foam reflectance in the near-infrared
and proposed that the foam reflectance can not be decoupled from
the refelectance by bubbles.  \cite{Stramski84a} concentrates on
light scattering by submerged bubbles in quiescent seas and shows
the scattering coefficient and the backscattering coefficient at
550 nm  in comparison with scattering and backscattering
coefficients of sea water as estimated from the chlorophyll-based
bio-optical models for Case 1 waters. In this exploratory paper,
we report on the influence of bubble clouds generated by breaking
waves on the remote sensing reflectance and calculate not only the
inherent optical, but also apparent optical properties using the
radiative transfer model. We show that the optical effects of
bubbles on remote sensing of the ocean color are significant.
Furthermore, we present a global map of volume fraction of air in
water. This map, together with the parameterization of the
microphysical properties, shows the significance of bubble clouds
on the global albedo of incoming solar energy. By proxy, we show
the influence of the bubble clouds on the remote sensing retrieval
of organic and inorganic components of the natural waters. It is
worth mentioning that the bubble clouds coincide with the upper
range of the euphotic zone and will, therefore, contribute to the
dynamics of the upper-ocean boundary layer, heat distribution, and
sea surface temperature \citep{Thorpe92a}. In fact, our initial
motivation for this work was an observation that the asymptotic
radiance distribution is established  close to the ocean surface
in apparent contradiction with  theoretical studies
\citep{Flatau99a}. Thus, the light field must become diffuse at
shallower depths than usually modeled. This leads to search for
alternative mechanisms influencing the light distribution. Thus,
the importance to light scattering  of the bubble clouds goes
beyond the remote sensing issues considered in this work.

In the next section, we discuss in more detail the  microphysical
and morphological  properties of bubble clouds, because they  have
a direct bearing on their optical properties and radiative
transfer.

\section{Physical properties of bubble clouds}

\subsection{Morphology of bubble clouds in natural waters}
\begin{figure}[h]
\btdplot{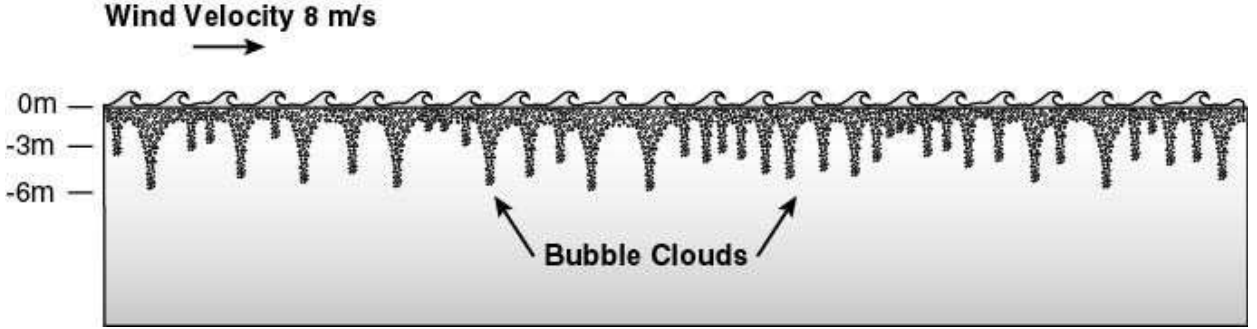}{1.} \caption{\label{bubblesc}
Bubble-stratocumulus (b-Sc) layer generated by breaking waves with
inhomogeneous deeper intrusions. The depth of the cloud is related
to the maximum significant wave height. }
\end{figure}

Individual bubble clouds are generated by breaking waves, persist
for several minutes \citep{Thorpe95a}, and reach to mean depths of
about $4H_s$, where $H_s$ is the significant wave height, but with
some clouds extending to about $6 H_s$.  There is evidence that at
high wind speeds, separate bubble clouds near the surface
coalesce, producing a stratus layer \citep{Thorpe95a}.
Fig~\ref{bubblesc} is based on a sonograph of \citet{Thorpe84a}.
The ``bubble-stratocumulus'' (b-Sc) is often observed by acoustic
means \citep{Farmer84a,Thorpe95a}. The depth of the b-Sc layer is
related to the wind speed and wind variability, but more
specifically it is set by larger waves,  such us  those breaking
predominantly in groups \citep{Thorpe95a}. The ``stratus layer''
description should not be taken too literally. For sufficiently
high winds there will be significant concentrations through out
the upper layer, but the variability within this layer can be very
high.

\subsection{\label{sec:bubble} Bubble cloud climatology} According
to \citet{Thorpe92a} in the absence of precipitation and in wind
speeds exceeding about 3 m~s$^{-1}$, wave breaking generally
provides the dominant source of bubbles.  The wind speed $W_{10}$
at 10 m above the mean sea surface level is used to parameterize
the volume fraction of air in water, $f= V_{\rm air} / V_w$, where
$V_{\rm air}$ is the volume of air, and $V_w$ is the volume of
water.  This parameterization is an approximation of a  more
complex wind-wave relationship.
\begin{figure}
\btdplot{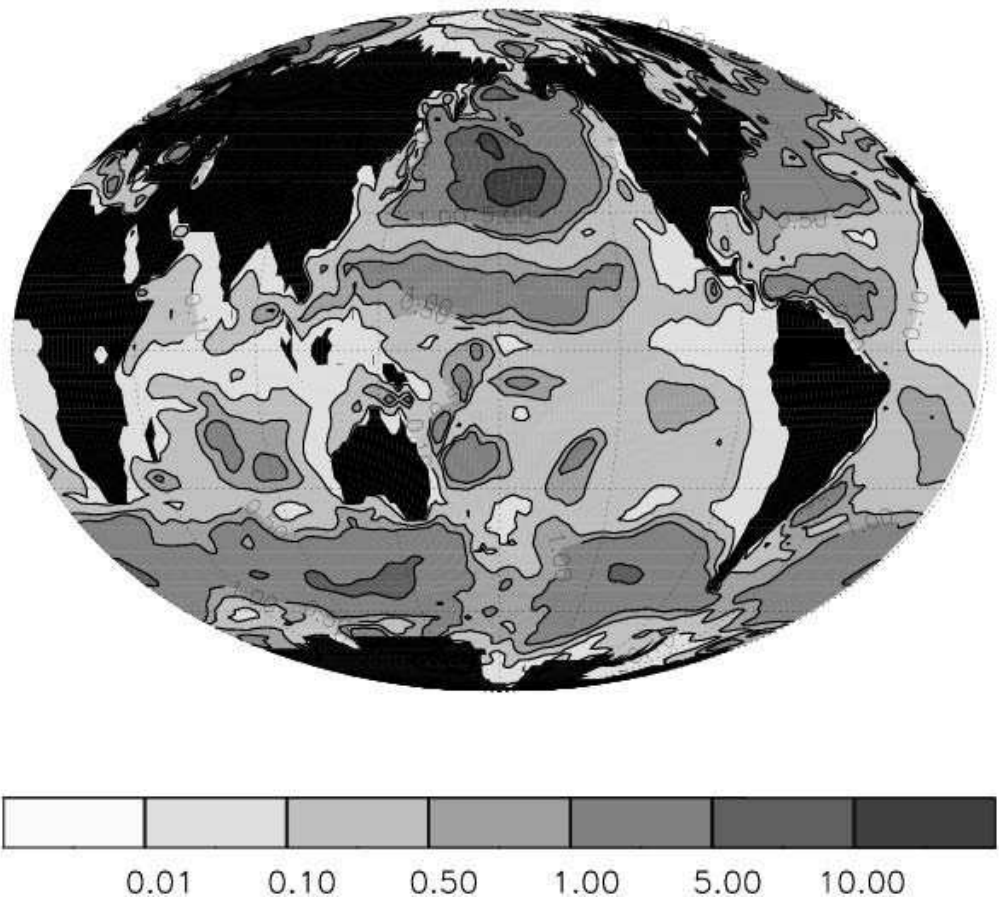}{1.} \caption{\label{janbub} January 1992,
monthly averaged volume fraction. Volume fraction $\times 10^6$.}
\end{figure}
Figure~\ref{janbub} shows the volume fraction of air in water estimated from the
$W_{10}$ winds for January of 1992.  The volume fraction and the wind speed are
assumed to follow a non-linear relationship \citep{Walsh87a}
\begin{equation}
f = f_0 W_{10}^{4.4} + f_1 \label{eqf}
\end{equation}
The coefficients $f_0$ and $f_1$ were calculated at 2m assuming that f= $10^{-8}$
for $W_{10}=6.2 {\rm m \; s^{-1}}$ and f= $10^{-7}$ for $W_{10}=10.5 {\rm m \;
s^{-1}}$ \citep{Walsh87a}.  \citet{Vagle92a} show that the volume fraction
decreases with depth, changing from about $10^{-6}$ at 0.3 m to $10^{-7}$ at 2.7
m.  We base our parameterization on these findings and extrapolate (\ref{eqf}) to
near-sea-surface depth using exponential fit.

The monthly averages of $f$ were obtained by employing 1992 daily surface winds
from the NCAR/NCEP \citep{Kalnay96a_short} reanalysis project, and averaging the
daily volume fractions for each month. Thus, Figure~\ref{janbub} is based on the
variability of the wind field on the scale of one day.  In the winter of the
northern hemisphere, one can observe maxima associated with the midlatitude storm
tracks in the Northern Pacific. Cyclogenesis, common in western parts of the
oceans during the winter, contributes to mixing and a large bubble cloud volume
fraction.  This can be observed to the east of the North American continent.  The
Intertropical Convergence Zone (ITCZ) region, with its associated deep
convection, may also be a region of enhanced production of bubbles. The winds of
the Southern Ocean have a strong effect on bubble formation during both summer
and winter.  In subtropical regions, to the west of the continents, the
subsidence associated with the descending branch of the Hadley circulation is
responsible for the relative minimum in $f$. It should be stressed that these
results are qualitative and that they can be improved by more detailed breaking
wave climatology models \citep{Kraus94a}.  The data in Fig~\ref{janbub} is
indicative of regions where bubble clouds are potentially important in the
interpretation of remotely sensed reflectance.

\subsection{Optical thickness} The size distribution of bubble
clouds determines the optical thickness and is, therefore, one of
the most critical parameters entering the theory. There are two
assumptions which simplify the development here: (a) we consider
light scattering in the geometric optics regime for which the
ratio proportional to bubble radius to wavelength $x= 2 \pi r /
\lambda$ is large and (b) we assume that there is an effective
radius $r_{\rm eff}$ which determines the optical properties of
the size distribution.  Both (a) and (b) are quite probable.  The
size parameter, $x = 50$, corresponding to a bubble radius of
approximately  5 ${\rm \mu m}$, is already in the geometric optics
regime, and $r_{\rm eff} = 10 {\rm \mu m}$ will satisfy (a).  For
bubbles with an effective radius of $r_{\rm eff}$, the volume
attenuation (equal to scattering for a non-absorbing sphere) can
be expressed as
\begin{equation}
b= Q_{\rm sca} {N \over V_w} s = 2 {V_{\rm air} \over V_w} {N \over V_{\rm air}}
s = 2 f {s \over v}.
\end{equation}
Thus
\begin{equation}
b= 2 {f \over r_{\rm eff}} \label{eqc}
\end{equation}
where $s=\pi r^2$ is the cross-section of a bubble with radius
$r$, $v$ is the volume of such a bubble. $N$ is the number of
bubbles in volume $V_w$ of water and  $f$ is the fraction of air
in a volume of water. The effective radius is defined as $r_{\rm
eff} = v/s$ \citep{Stephens90a,King93a,Bricaud86a}. Scattering
efficiency defines how much of incoming light is being ``blocked''
by a particle by scattering processes, for large size parameters
$Q_{\rm sca}$ tends to 2. This issue is discussed in detail by
\cite{Bohren83a}.

The physical significance of $f$ comes from the fact that it is
determined by the large scale forcing such as the wind field.
Thus, for a given synoptic or climatological setting, the mixing
ratio $f$ is, to some extent, pre-determined. On the other hand,
the effective radius $r_{\rm eff}$ depends on processes of much
smaller scale then the large scale. These processes are
coagulation, coalescence, coating by organic material, saturation,
buoyancy, pressure, etc. Thus, Eq.~\ref{eqc} defines the optical
properties of bubble clouds on both the large- and sub-scales. The
expression $b=2 f / r_{\rm eff}$ holds for polydispersions, and
the only difference with a monodispersion is that $r_{\rm eff}$ is
defined via distribution averaged $s$ and $v$.

\subsection{Effective radius of the size distribution} Numerous
observations of bubble size distributions are reported in the
literature based on acoustic, photographic, optical, and
holographic methods \citep{Wu88a1}. \cite{Akulichev87a} summarize
results from 22 experiments using different techniques. As  the
origins of bubbles are biological (within the volume and at the
bottom), as well as physical (at the rough surface), we may expect
large regional and temporal variations of bubble concentration
between coastal and open oceanic waters and between plankton bloom
or no-bloom conditions \citep{Thorpe92a}.

\begin{figure}
\btdplot{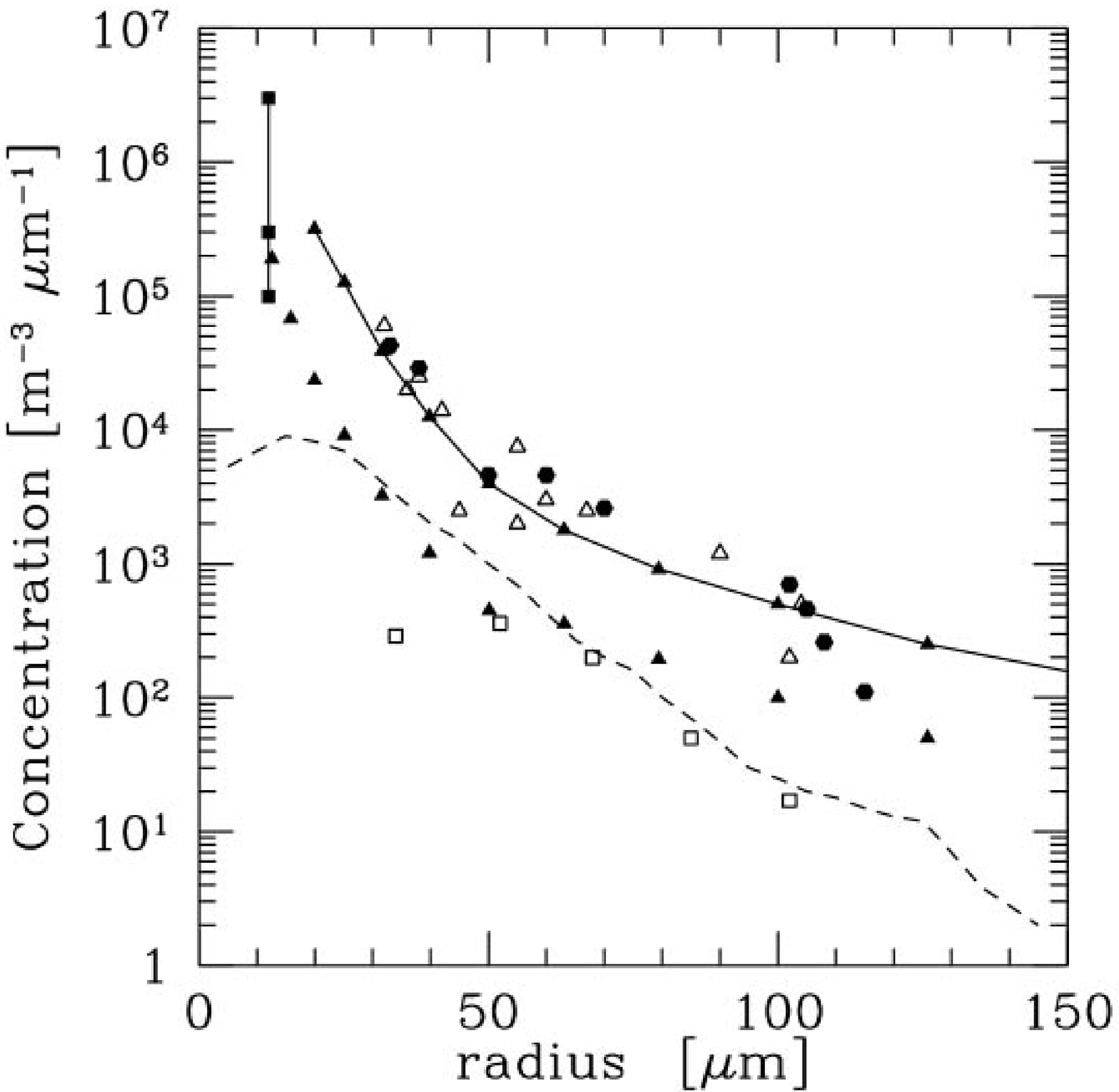}{1.} \caption{\label{concentration}
Comparison of bubble spectra under breaking waves and quiescent
sea. The solid squares connected with the vertical solid line are
from laser holographic data \cite{OHern88a}. Open triangles are
from \cite{Baldy88a} and include data at 30cm depth with wind and
swell, and at 25cm with wind only. The solid hexagons are data
based on \cite{Medwin89a} obtained in the open sea at 25cm depth
under the water surface during winds of 12${\rm m \; s}^{-1}$. The
solid triangles joined by the solid line  and solid triangles are
average bubble densities measured under comparable conditions at
different seasons. The solid triangles joined by solid lines are
concentrations at 4m depth, 3.3${\rm  m \; s}^{-1}$ wind speed,
obtained on August 7, 1975 in Monterey Bay. The solid triangles
are for midafternoon, February 10-16, 1965 at Mission Bay, San
Diego, 3m below the surface, 1.7-2.8${\rm  m\; s}{-1}$ winds. The
dashed line is from a numerical model \citep{Thorpe92a}. The mean
concentration in the steady-state model are plotted at temperature
0C. The open squares show \citep{Johnson79a} observations at 4m in
wind speed 11-13${\rm m \; s}^{-1}$. }
\end{figure}

Figure~\ref{concentration} presents a comparison of bubble spectra
under breaking waves and quiescent sea. Recently published
observations, using laser holography near the ocean surface, have
shown that the densities of 10 to 15$\mu$m radius bubbles can be
as high as $10^6$ (per cubic meter per micron radius increment)
within 3 m of the surface of quiescent seas \citep{OHern88a}.
These results are plotted as solid squares connected with a
vertical solid line. The majority of bubbles injected into the
surface layers of natural waters is unstable, either dissolving
due to enhanced surface tension and hydrostatic pressures or
rising to the air-water interface where the bubbles  break
\citep{Johnson87a}. However, bubbles with long residence times,
i.e. stable microbubbles, have been observed. For example
\cite{Medwin77a} observed nearly $2.5 \times 10^6$of  bubbles per
cubic meter in the radius range $18-355$ $\mu$m for small wind
speeds. One of the stabilization mechanisms
\citep{Mulhearn81a,Johnson87a} assumes that the surfactant
material is a natural degradation product of chlorophyll, present
in almost all photosynthesizing algae. \cite{Isao90a} have
observed very large populations of neutrally buoyant particles
with radii between $0.1-1 \mu$m. \cite{Johnson87a} and
\cite{Thorpe92a} proposed another stabilization mechanism based on
monolayers of adsorbed particles. Numerical modeling
\citep{Thorpe92a} can be used to study the effects of water
temperature, dissolved gas saturation levels, and particulate
concentrations on the size distribution of subsurface bubbles. The
results of such numerical models provide additional evidence for
the existence of a small size bubble fraction which is not
adequately measured by acoustic or photographic techniques. The
dashed line on Fig.~\ref{concentration} presents the mean
concentration in the model steady-state for the water temperature
of 0C.  It can be seen that maximum concentration is around
$15\mu$m and it is 2.5 orders of magnitude larger than for
100$\mu$m bubbles. Other results presented
(Figure~\ref{concentration}) are those of \cite{Johnson79a}
observations at 4m in wind speed 11-13${\rm m \; s}^{-1}$ (open
squares). In situ acoustic measurements of microbubbles at sea by
\cite{Medwin77a} are plotted as solid triangles. The solid
triangles joined by a solid line are concentrations at 4m depth
and  3.3m$s^{-1}$ wind speed. These spectra were obtained on
August 7, 1975 in Monterey Bay. The solid triangles are for
midafternoon, February 10-16, 1965 at Mission Bay, San Diego, 3m
below the surface, and in 1.7-2.8${\rm m \; s}^{-1}$ winds. The
open triangles are from \cite{Baldy88a} and include data at 30cm
depth with wind and swell and at 25cm with wind only. They are
based on extensive laboratory experiments. The solid hexagons are
data from \cite{Medwin89a}  acoustic measurements obtained in the
open sea at 25cm depth under the water surface during 12${\rm
m\;s}^{-1}$ winds under spilling breakers. The literature reviewed
here and encapsulated in Figure~\ref{concentration} is a mixture
of descriptions of the effect of active wave breaking, and of
stabilised microbubbles observed largely in coastal situations.
Currently, it is not clear how to parameterise the stabilized
bubbles.
From results such as those presented in Fig.~\ref{concentration}
in the case of ``transient,'' open ocean bubbles, it can be
estimated that the size distribution follows the power law
dependence $n(r) \propto r^{-a}$ and $a \approx 4$
\citep{Walsh87a, Wu88a1}. Even though small microbubbles may not
contribute to the total mass, they may be important for the light
scattering. Therefore, it is of interest to estimate the
contribution  of small bubbles to  the optical thickness. Assuming
that microbubbles are spherical ($V_{\rm air} = 4/3 \pi r^3 N$) we
can show that the optical thickness ($\tau = b h$)  of a layer
with geometrical thickness $h$ is
\begin{equation}
b \propto h  V_{\rm air}^{2/3} N^{1/3} \label{eqd}
\end{equation}
Both $h$ and $V_{\rm air}$ are assumed to be fixed. It can be seen
from Eq.~\ref{eqd} that the contribution to optical thickness by
very small particles will be the same as that by very large
particles if their concentration varies as $N^3$ (or steeper).
Indeed is the case  \citep{Walsh87a, Wu88a1}.

What remains to be defined is the effective radius $r_{\rm eff}$.
On the basis of measurements, estimates of small particle
fraction, existence of background 1$\mu$m microbubbles, modeling
predictions, and steep slope of microbubble size distributions, we
decided to use $r_{\rm eff} = 10 {\rm \mu m}$ as a typical
``radiative response radius.'' This choice does not exclude
existence of larger or smaller particles. The real value may be
between $1\mu$m for stabilized particles and $50\mu$m for the open
ocean and will depend on many environmental factors such as storm
passage, wind speed, swell, wind variability, phytoplankton
concentration, water temperature, gas saturation, and other
properties. The 10-15 fold increase in the size of effective
radius or similar decrease in the air volume fraction will reduce
the importance of air bubbles to a very small effect. It may be
instructive to calculate the scattering coefficient $b$ for
typical size distribution of bubbles in water. In such case we
have
\begin{equation}
b= \int Q_{\rm sca} \pi r^2 {dN(r) \over V_w}
\end{equation}
or $b=2f s/v$ where $s= \int \pi r^2 dN(r) $ and $v=\int 4/3 \pi r^3 dN(r)$, and
$d N(r)$ is the number of bubbles between $r$ and $r + dr$ in a volume of water
$V_w$. For typical size distribution of bubbles in water $dN(r)/dr \propto 1/r^4$
we have
\begin{equation}
r_{\rm eff} = {4 \over 3} \ln(r_1/r_0) / \left( {{1 \over r_0} - {1 \over r_1}}
\right)
\end{equation}
Consider $r_1 = 150$ and $r_0 = 10$ micrometers. This gives
$r_{\rm eff} \sim 3.6 r_0$ which shows that the choice of small
bubble cut-off is important for the bubbles' optical properties.
However, the choice of this cutoff is non-trivial because the
spectrum of small bubbles is not understood well at present.

We close this section with some general comment about the
effective radius. It is not a directly measurable quantity and, in
essence, it defines how dispersed the given amount of mass is.
Scattering of incoming solar radiation  is sensitive to total
projected surface rather than to total mass. For this reason the
effective radius is commonly used in radiation calculations.
However, it should be stressed that the effective radius is a
semi-inherent optical property because it carries information not
only about the size itself, but also about the orientation of
particles, their morphology, coating, size distribution, or
departure from a spherical shape. In addition, estimates  of
effective radius, as used in satellite remote sensing, often
contain bias due to unrealistic assumptions about other optical
properties such as optical thickness, leakage of photons due to
horizontal transfer, wavelengths,  or technique employed in
retrieval. In that sense the  effective radius is also used (or
abused) as a semi-apparent optical property.

\section{Results}

\subsection{Numerical model} The numerical radiative transfer model
used in this study is a slightly modified version of the
Hydrolight 3.0 code \citep{Mobley94a,Mobley94b_short}.  In brief,
this model computes from first principles the radiance
distribution within, and leaving, any plane-parallel water body.
Input to the model consists of the absorbing and scattering
properties of the water body, the nature of the wind-blown sea
surface and of the bottom of the water column, and the sun and sky
radiance incident on the sea surface.

\begin{figure}
\btdplot{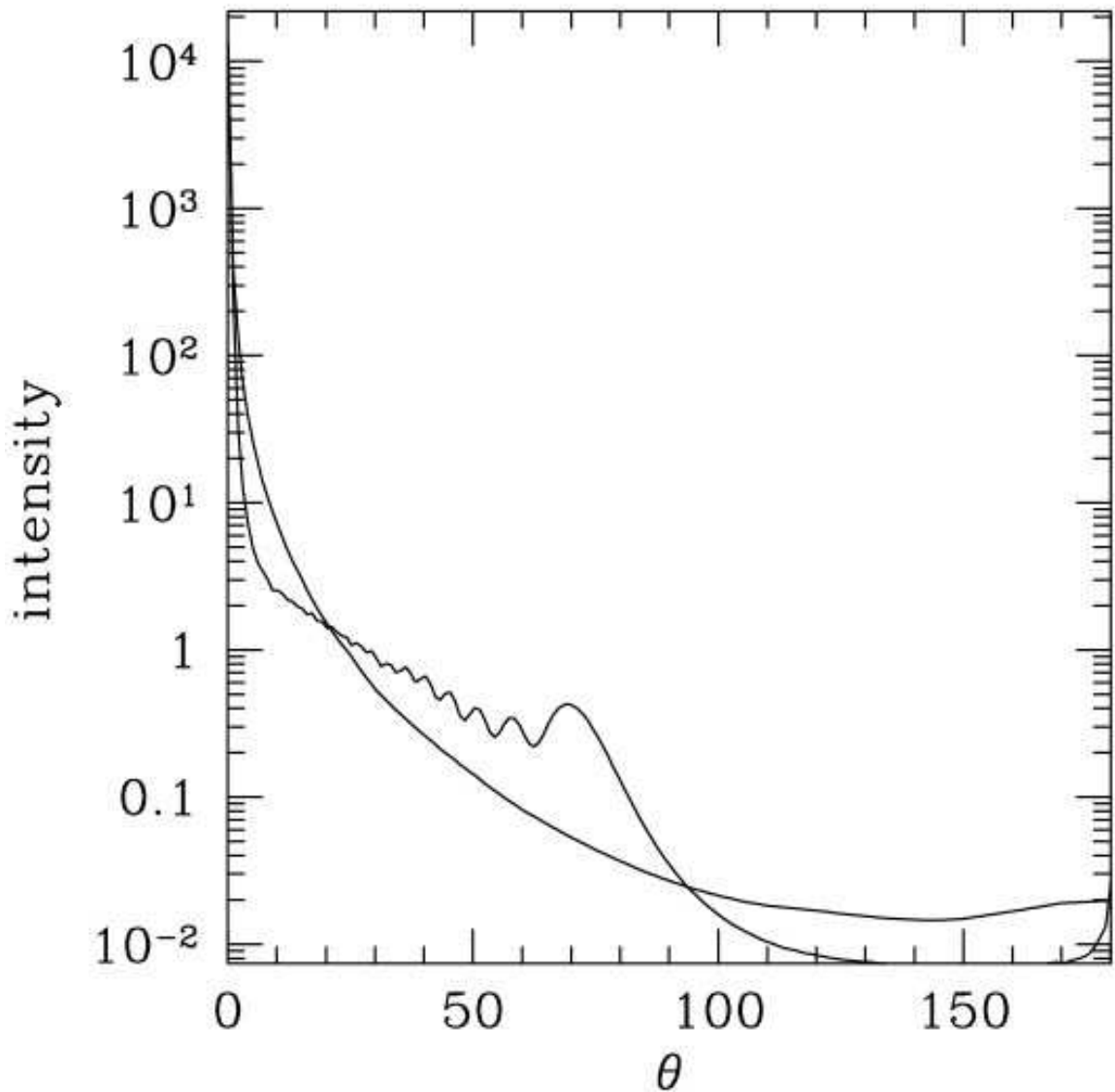}{1.} \caption{\label{comp} Unpolarized phase
function for a uniform distribution of bubbles between size
parameter 100 and 300 and refractive index $m=3/4$ as a function
of scattering angle $\theta$. Also shown is Petzold's phase
function. The bubble phase function has peak at 70 degrees.}
\end{figure}

Pure sea water absorption and scattering coefficients are determined from the
data of \citet{Pope97a}.  35 model wavebands were specified to cover the 400-700
nm region with a typical resolution of 10nm.  The water column was specified as
infinitely deep.  Up to 62 depth layers, extending to 50 meters, were specified
with a resolution of 0.5 m close to the surface.  A clear sky was assumed, but
the diffuse sky radiance was included.  Three or four component systems were
considered, consisting of pure water, particulates  with or without bubble
clouds, and dissolved organic matter.  The spectral absorption of dissolved
organic matter was defined as
\begin{equation}
a(\lambda) =a(\lambda_0) \exp[-0.014(\lambda-\lambda_0)]
\end{equation}
where $a(\lambda_0)=0.1$ ${\rm m^{-1}}$, $\lambda_0=440$ nm \citep{Bricaud81a}.
The phase function of phytoplankton was defined as an average of Petzold's clear
ocean, coastal ocean, and turbid harbor cases \citep{Mobley94a,Tyler77a}.  The
bubble cloud phase function was calculated for the real relative refractive index
$m=0.75$ as an average for the size parameter range between $x=100-300$ with a
resolution $\Delta x = 1$ using Wiscombe \citep{Fiedler-Ferrari91a} code.
Figure~\ref{comp} compares Petzold and bubble phase functions.
The main reason for the size distribution average was to remove
transient spikes. This normalized phase function is scaled by $b$
calculated from Eq.~\ref{eqc}. The volume scattering and
absorption coefficients for particulates were determined from the
Gordon-Morel model \citep{Mobley94a,Gordon83a,Morel88a}
\begin{equation}
b_p={550 \over \lambda} 0.3 C^{0.62}
\end{equation}
\begin{equation}
a^{\rm Case 1}(\lambda)=[a_w(\lambda)+0.06 a_c^\star (\lambda) C^{0.65}] [1+0.2
\exp(-0.014(\lambda-440))]
\end{equation}
Here $a_w(\lambda)$ is the absorption coefficient of pure water, $a_c^\star$ is
the chlorophyll-specific absorption coefficient, and $C$ is the chlorophyll
concentration in ${\rm mg\; m^{-3}}$ \citep{Mobley94a}.  The chlorophyll
concentration was set as constant (well-mixed) with depth and equal to 0.8${\rm
mg\; m^{-3}}$ or 0.08${\rm mg\; m^{-3}}$.  We used Case 1 water parameterization
but assumed that not all dissolved organic material  is correlated with the
chlorophyll concentration. In an apparent contradiction, the wind speed which
defined surface reflectance and transmittance functions due to the wind-blown
water surface was set to 0. The reason for this was  to estimate the effect on
remote sensing properties of the sub-surface bubble clouds. However, we
investigated the sensitivity of the reflectance to change in the wind speed
between 0 and 10 ${\rm m s^{-1}}$, and the effect was small compared to the
influence of bubbles.  The azimuth direction was divided into 24 equally spaced
sectors, the zenith-nadir range was divided into 20 equally spaced sectors.  The
profile of the bubble cloud volume fraction was determined by the expression
\begin{equation}
f(z)=f_0 (f_1/f_0)^p
\end{equation}
where $f_0=f(z_0)=10^{-6}$, $f_1=f(z_1)=10^{-7}$, $z_0=0$ m, $z_1=8$ m, and
$p=(z-z_0)/(z_1-z_0)$.  Our choice of the speed of attenuation is perhaps
unreasonably gradual, except for the case of stable microbubbles. On the other
hand this choice is of secondary importance for the remote sensing properties of
bubble clouds which are dominated by the surface volume fraction.  In-water
asymptotic light field is discussed by \cite{Flatau99a}. The scattering was
conservative. The volume attenuation was determined from the asymptotic
expression (\ref{eqc}). The probability distribution function for scattering was
discretized and was set as constant throughout the depth range. We calculated the
asymmetry parameter to be $g \approx 0.85$. This parameter defines the
probability of photon scattering towards the forward or backward hemisphere. It
is equal to 1 if  all photons are scattered forward and to -1 if all photons are
scattered backward.

\subsection{Remote sensing reflectance} 
\begin{figure}
\btdplot{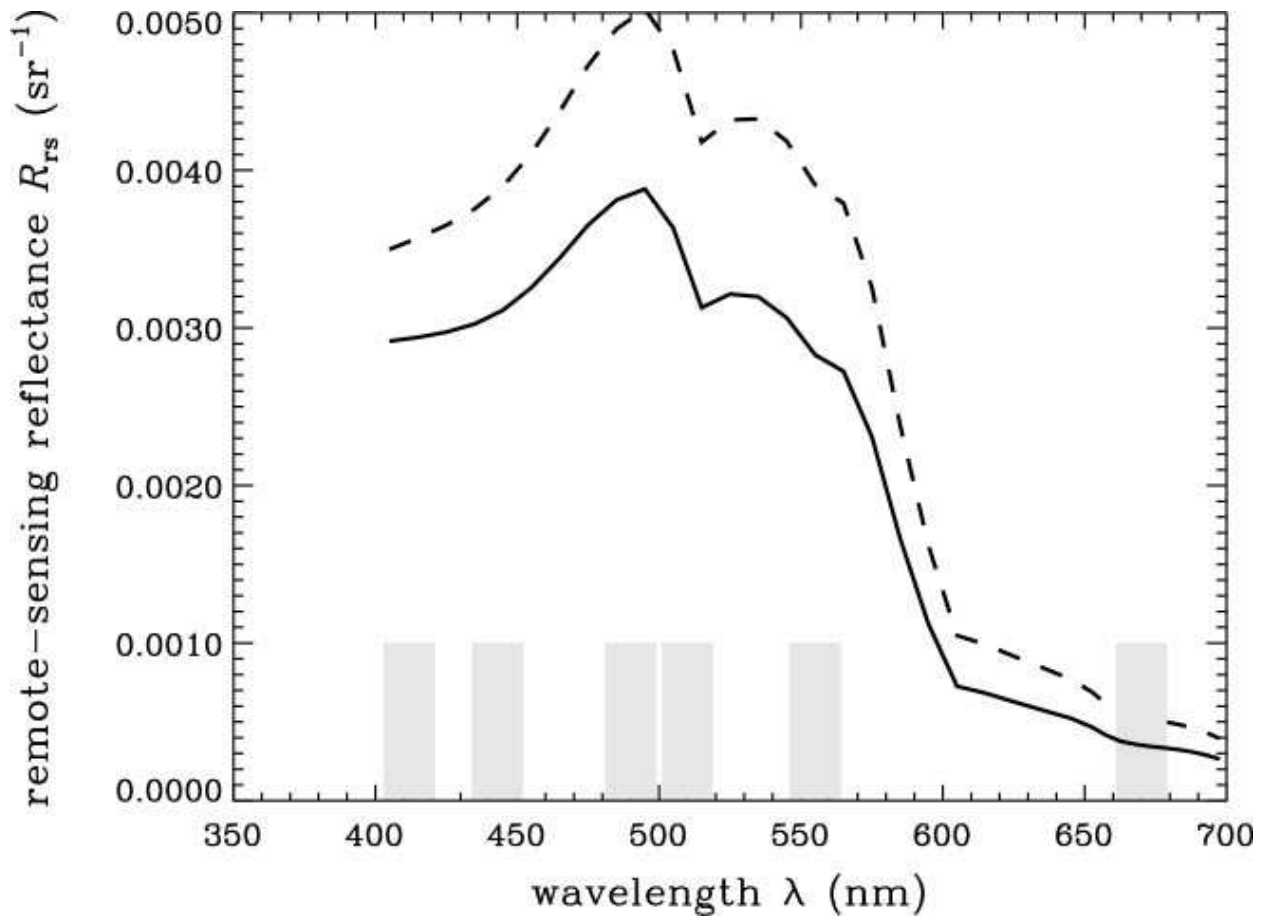}{1.} \caption{\label{rrs} The solid line is
the remote-sensing reflectance for the 3-component system composed
of water, DOM, and particulates (no microbubbles) and the dashed
line is for the 4-component system (microbubbles included). Same
chlorophyll concentration (0.8${\rm mg\; m^{-3}}$) in both cases.
Effective radius $r_{\rm eff}=10\mu m$. Hydrolight run with 62
layers, maximum depth 50m, maximum bubble depth 8m, 35 wavenumbers
between 400-700nm, sun zenith angle 50. The grey rectangles
indicate SeaWiFS (ocean color satellite) wavelengths.}
\end{figure}

\begin{figure}
\btdplot{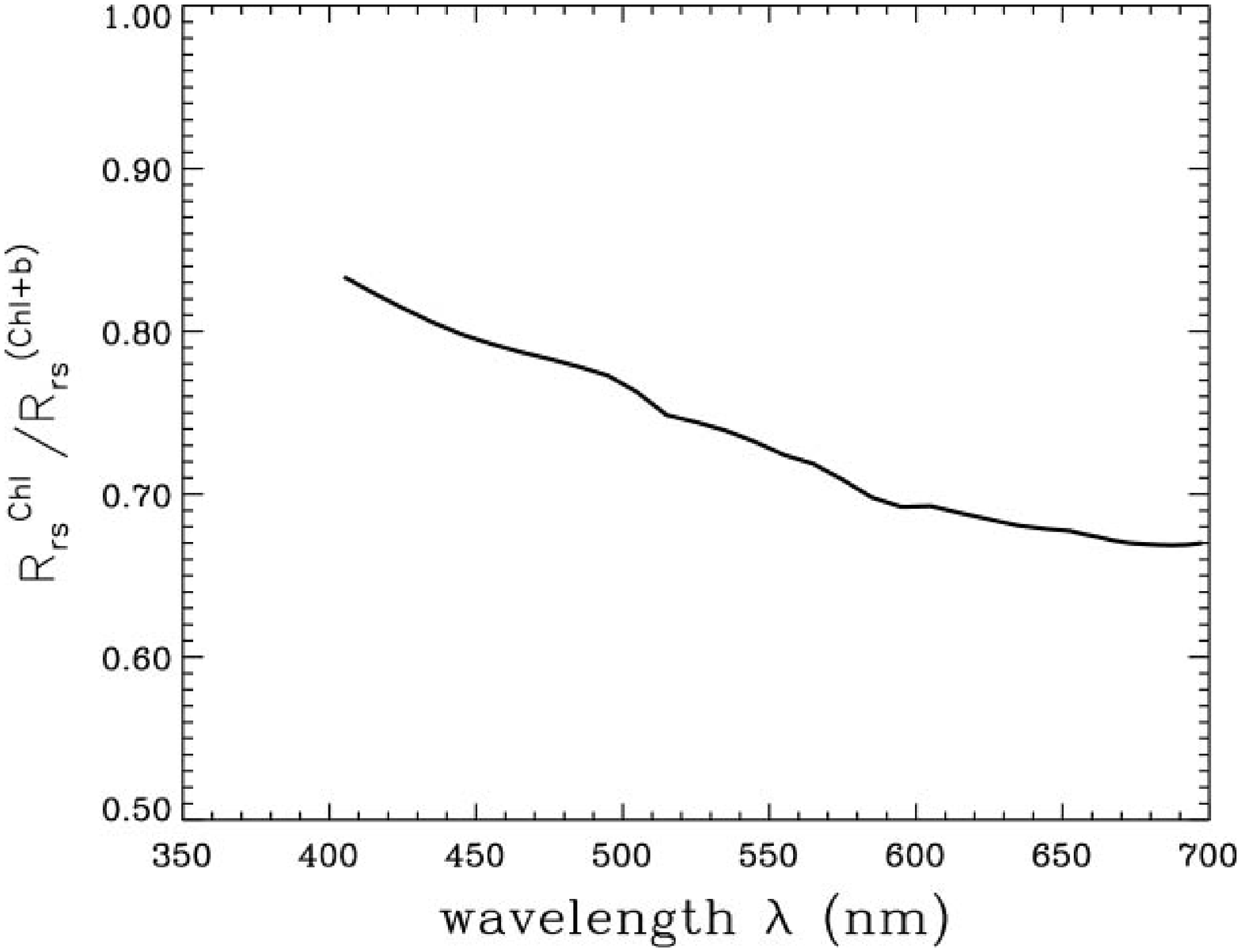}{1.} \caption{\label{ratiorrs} The ratio of
remote-sensing reflectances ${\rm ratio}(\lambda)=R_{\rm rs}^{\rm
Chl}/R_{\rm rs}^{\rm Chl+b}$. Same experiment as in
Fig.~\ref{rrs}. }
\end{figure}

The reference model runs
were performed with the 3 component system of phytoplankton, pure
water, and DOM with a constant chlorophyll profile. Two cases were
computed (but only one is presented), representative of clear
coastal ($C=0.8$) and oceanic ($C=0.08$) water.  The remote sensed
reflectance is defined as
\begin{equation}
R_{\rm rs} (\lambda) = {L_w(\lambda) \over E_d (\lambda)}
\end{equation}
where $E_d$ is the downwelling irradiance onto the sea surface,
and $L_w(\lambda)$ is the upwelling water-leaving radiance.  The
remote-sensing radiance is a measure \citep{Mobley94a} of how much
of the downwelling light that is incident onto the water surface
is returned into the zenith direction.  The remote sensing
reflectances are plotted in Fig.~\ref{rrs}.  The physical
importance of the remote-sensing reflectance is evident from
asymptotic theories which relate $R_{\rm rs}$ to inherent optical
properties \citep{Zaneveld95a}
\begin{equation}
R_{\rm rs} \propto {\beta(\pi-\theta_m) \over a}
\end{equation}
where $\beta$ is the phase function, $\theta_m$ is related to the sun zenith
angle, and $a$ is the absorption coefficient. Thus, $R_{\rm rs}$ is approximately
proportional to the  probability of back- or side-scattering, and inversly
proportional to the absorption of water column. Figure~\ref{rrs} shows the
remote-sensing reflectance for the 3- and 4-component system with and without
ocean bubbles but with constant pigment amount.  The total single scattering
albedo is strongly influenced by scattering from air bubbles. This leads to
enhanced reflectance at all wavelengths. The results for $C=0.08$ (not presented)
show even larger sensitivity. In Fig~\ref{rrs} the gray rectangles indicate bands
(wavelenghts)  which are used by the current ocean color satellite instrument
(SeaWiFS). It is of interest to comment on the remote sensing of pigments and
bubble cloud retrievals. Consider an algorithm based on the ratio of
remote-sensing reflectance and define the ratio of remote-sensing reflectances
without (Chl) and with (Chl+b) microbubbles as
\begin{equation}
{\rm ratio}(\lambda)=R_{\rm rs}^{\rm Chl}/R_{\rm rs}^{\rm Chl+b}.
\end{equation}
Figure~\ref{ratiorrs} shows ${\rm ratio}(\lambda)$.  Performance
of the pigment algorithms based on the ratio of reflectances will
depend on ${\rm ratio}(\lambda_1)/{\rm ratio}(\lambda_2)$.
Submerged microbubble clouds seem to be wavelength-selective and
even the ratio algorithms may require slight systematic
correction.  Given the increased sensitivity of the current
generation of ocean color instruments, the absolute value of the
radiances at the top of the atmosphere can be used for pigment
retrievals.

\section{Summary}

Our calculations indicate that the optical effects of submerged
microbubbles on the remote sensing reflectance of the ocean color
are significant.  These results are of importance for the
retrievals of pigments from the ocean color measurements and for
studies of the energetics of the ocean mixed layer.  We provide
information on how to reduce the systematic error due to
microbubbles in pigment retrieval schemes via the ${\rm
ratio}(\lambda)$. We also derive apparent optical property of
bubbles - remote sensing reflectance - for the whole solar
spectrum. This AOP is directly observable by the satellites and
remote sensors. We expect that these and similar AOPs will have to
be invoked in the case of hyperspectral retrievals for Case 2
waters where the signals from minerals, bubbles, chlorophyll, and
dissolved organic material (CDOM)  are not well correlated. New
algorithms for current satellite instruments such as MODIS and
SeaWIFS should employ this information.

We also present a global map of the volume fraction of air in
water derived from daily wind speed data. We expect that such a
map can be improved by knowing the day-to-day variability of the
wind-wave relationship and better estimates of the volume
fraction.

The paper is exploratory. Therefore, it is perhaps worth playing
{\it advocatus diaboli} and speculate why the bubble clouds may
not be important,  at least in current satellite ocean color
retrieval practice.  Here are some reasons: (1) The high wind and
clouds are correlated. This masks (bias) the effect of bubbles as
observed from the satellites; (2) Both whitecaps
\citep{Wu88a,Gordon94a,Frouin96a} and bubble clouds are correlated
via their dependence on wind speed. Therefore, our results, as
well as the hypothesis of \citet{Frouin96a}, indicate that
reflectance of foam has to be considered together with the
reflectance due to bubble clouds. On the other hand, there are
cases in which strong wind is not correlated with clouds. For
example, the cross equatorial flow during the summer monsoon in
the southern Indian Ocean is strong but the ITCZ position is in
the northern hemisphere.

It is interesting to note that stabilized, coated microbubbles are hypothesized
to be correlated to phytoplankton and CDOM concentrations; we need
parameterization of this process.  The optical properties of the first several
meters below the surface are difficult to measure and are often removed from data
due to experimental problems such as ship shadow or wave activity.  This is the
region where more detailed studies are needed.

\section*{Acknowledgements} P. J. Flatau was supported in part by the
Office of Naval Research Young Investigator Program and NASA
SIMBIOS program.  M. Flatau acknowledges NOAA/UCAR Global Climate
Change Fellowship and  J. R. V. Zaneveld acknowledges support of
the Environmental Optics program of the Office of Naval Research
and the Biogeochemistry program of NASA.  C. D. Mobley
acknowledges support of the Environmental Optics program of the
Office of Naval Research, which also supported in part the
development of the Hydrolight model.

\section*{References}
\bibliography{local}
\bibliographystyle{jas99}

\end{document}